# Visualizing the interplay of dual electronic nematicities in kagome superconductors

Yunmei Zhang[1#], Jun Zhan[2#], Ping Wu[1], Yun-Peng Huang[3], Qixiao Yuan[1], Hongyu Li[4], Zhuying Wang[1], Wanru Ma[1,5], Shuikang Yu[1], Kunming Zhang[1], Wanlin Cheng[1], Deshu Chen[1], Minrui Chen[1], Tao Wu[4,5], Ziji Xiang[4,5], Xianxin Wu[3*], Zhenyu Wang[1,5*], and Xianhui Chen[1,4,5*]

[1]Department of Physics, University of Science and Technology of China, Hefei, Anhui 230026, China

[2] Beijing National Laboratory for Condensed Matter Physics and Institute of Physics, Chinese Academy of Sciences, Beijing 100190, China

[3]CAS Key Laboratory of Theoretical Physics, Institute of Theoretical Physics, Chinese Academy of Sciences, Beijing 100190, China

[4] Hefei National Research Center for Physical Sciences at the Microscale, University of Science and Technology of China, Hefei, 230026, China

[5] Hefei National Laboratory, University of Science and Technology of China, Hefei, 230088, China

[#]These authors contributed equally to this work

*Correspondence and requests for materials should be addressed to X. W. (xxwu@itp.ac.cn), Z.W. (zywang2@ustc.edu.cn) or X.-H.C. (chenxh@ustc.edu.cn).

**Kagome superconductor $A$V$_3$Sb$_5$ ($A$ stands for K, Rb, and Cs) hosts a wealth of intertwined electronic orders driven by geometric frustration and electron correlations. Among them, the breaking of rotational and/or time-reversal symmetry, observed within the triple-$Q$ charge density wave (CDW) phase yet exhibiting a more complex temperature dependence, remains a central puzzle. Here, by using scanning tunneling microscopy to study the electronic structures of CsV$_3$Sb$_5$ as a function of temperature and Ti doping, we disentangle the interrelation between two distinct nematic order parameters, one associated with the CDW and the other manifested as C$_2$ distortion of the V-$d_{x^2-y^2}$ Fermi pockets without breaking transition symmetry. The latter persists to high doping levels and high temperatures where the long-range CDW is fully suppressed. Moreover, its nematic director is oriented in a lattice direction distinct from that of the CDW-induced nematicity at intermediate doping, and eventually aligns with the strong nematic CDW order in the pristine compound where the quasiparticles of vanadium orbitals become coherent below a lower characteristic temperature. These observations, combined with Ginzburg–Landau analysis, reveal a rich interplay between two nematic orders that can be assigned to distinct kagome-lattice orbitals. Our results shed new light on the enigmatic intertwined orders in this family and establish a rare material platform in which dual nematic orders coexist and couple to give rise to unusual correlated phenomena.**

In strongly correlated quantum materials, multiple electronic orders with distinct broken symmetries that are nearly degenerate in energy often emerge in a coupled manner, leading to complex emergent phenomena. This intriguing set of electronic orders can be described in terms of intertwined orders (1–3), as initially proposed for high-temperature superconductors, to emphasize their cooperative interrelation beyond mere competition. Recently, the landscape of intertwined orders has become remarkably rich in systems subject to strong geometrical frustration (4–6). One example is the kagome lattice (7,8), where the interplay of sublattice texture and geometric frustration leads to a plethora of electronic orders even in the presence of relatively weak electron–electron interactions (9–12).

The layered kagome superconductors $A$V$_3$Sb$_5$ have emerged as compelling platforms for investigating intertwined orders on a frustrated lattice (13–17). A key feature of the band structure here is the presence of multiple types of van Hove singularities (VHSs) arising from different vanadium $d$-orbitals located near the Fermi level (18–20), which can potentially promote a variety of symmetry-broken phases (21–29). This material family first undergoes a $2 \times 2 \times 2$ charge density wave (CDW) transition at T$_{CDW}$ ≈ 80 -100 K (17), establishing an intriguing backdrop for a cascade of intertwined symmetry-breaking orders, including possible time-reversal symmetry breaking (30–36), rotational symmetry breaking (37–42), and pair density waves (43,44).

In the prototypical compound CsV$_3$Sb$_5$, the interlayer stacking of the CDW order, characterized by a staggered tri-hexagonal configuration as illustrated in Fig. 1(a), breaks the rotational symmetry of the bulk lattice. However, several experimental probes have revealed an enigmatic change in electronic anisotropy below a lower temperature scale, T* ≈ 35 K (32,36,38,39,45–47), which has been attributed to electronic nematicity (38,39) or potential orbital current order (32,36,46). The microscopic nature of this low-temperature state is among the most compelling open questions. From a spectroscopic perspective, this state exhibits two prominent in-plane features: pronounced nematic CDW modulation (39,41) and a twofold (C$_2$) symmetric energy spectrum in momentum-space eigenstates (41,42,48); these two features appear to be locked to one preferred lattice direction (42,48). However, the relationship between them, i.e., whether they are consequences of the same ordering phenomenon or originate from distinct mechanisms, remains unknown. Furthermore, the electronic coherence of vanadium $d$-orbitals vanishes above T* (41,42), suggesting a possible hidden phase that intertwines intricately with the nematic CDW order. Therefore, disentangling the potential coexisting order parameters under external perturbations and elucidating their interplay constitute essential steps toward solving the rich puzzle.

Titanium substitution on the vanadium sites offers a direct means to tune the CDW order and investigate its interplay with other electronic states (49–52). In this work, we utilize spectroscopic-imaging scanning tunnelling microscopy (STM) to probe the local electronic structure in CsV$_{3-x}$Ti$_x$Sb$_5$ (x = 0.0, 0.12 and 0.18) and track its evolution as a function of doping and temperature. The main findings are summarized in Fig. 1(c). In samples with x = 0.18, where the CDW order is significantly suppressed, we identify a new intra-unit-cell ($\boldsymbol{q} = 0$) electronic state. This state manifests as a C$_2$-symmetric distortion of the V-$d_{x^2-y^2}$ Fermi pockets revealed by quasiparticle interference (QPI) mapping, reminiscent of the nematicity observed in the titanium-based kagome metal $A$Ti$_3$Bi$_5$ (53–55), and can persist at high temperatures. When this $q = 0$ nematic state coexists with the nematic CDW order, the two nematic directors can be either decoupled or locked together, depending on their interplay that varies with doping. On the basis of these observations, we introduce two nematic order parameters originating from different in-plane and out-of-plane vanadium orbitals and discuss their rich interplay within the framework of Ginzburg–Landau theory.

We first briefly introduce the QPI characteristics of CsV$_3$Sb$_5$ (37,42,52). A schematic of the constant-energy contours near the Fermi level is presented in Fig. 1(b). The dominant QPI channels include intra-band scattering involving the Sb-$p_z$ orbital ($q_0$; orange), scattering between parallel edges of the triangular V-$d_{x^2-y^2}$ pockets ($q_1$; blue), scattering connecting arcs of the same pockets ($q_2$; violet), and scattering connecting the opposite edge segments of the hexagonal V-$d_{yz}$ pocket ($q_3$; purple). Using high-precision QPI images that are obtained by Fourier-transforming differential conductance maps acquired over large fields of view (typically 80 nm × 80 nm), we then access the electronic structure, uncover potential symmetry-breaking tendencies, and track their evolution with temperature and doping.

We begin our investigation with the x=0.18 samples, where no signature of CDW transition is detected in bulk transport measurements (56; Fig. S1). One STM topography image acquired on the Sb-terminated surface is shown in Fig. 2(a). The dumbbell-shaped protrusions correspond to Ti substitutions in the underlying kagome layer, enabling a precise determination of the local Ti concentration through direct counting (56; Table S1). Furthermore, Fourier transforms (FTs) of topographic images recorded at various bias voltages, as exemplified here at 20 mV (Fig. 2(b)), reveal no detectable features at the $2 \times 2$ CDW wavevectors.

While bulk measurements and STM data consistently indicate that the long-range CDW order is fully suppressed at x=0.18, the QPI patterns unexpectedly reveal a pronounced anisotropy in the momentum-space electronic structure. Figure 2(d) presents the raw Fourier transform of a differential conductance map at low bias voltage (Fig. 2(c)). While the $q_0$ feature (associated with Sb-$p_z$ orbitals) appears isotropic, the wavevectors of the six petal-shaped scatterings ($q_1$ and $q_2$; arising from V-$d_{x^2-y^2}$ orbitals) exhibit striking C$_2$ symmetry. Specifically, the wavevector along the $\Gamma K_b$ direction is slightly larger than those along $\Gamma K_a$ and $\Gamma K_c$. This difference can be better visualized in the angular-dependent QPI intensity map as shown in Fig. 2(e): $q_0$ appears as an angle-independent straight line, whereas the wavevectors of $q_1$ and $q_2$ are larger at 90° and 270° ($\Gamma K_b$). The energy dispersions of these scattering wavevectors are displayed in Fig. 2(f), which further confirms that the $q_0$ wave vectors are nearly identical along all three directions, whereas the $q_1$ and $q_2$ wavevectors along the $b$ direction are systematically larger than those along the $a$ and $c$ directions.

This anisotropy in the electronic band structures has been consistently observed in all three measured x=0.18 samples (56; Fig. S2), and we attribute such anisotropy to an electronic origin beyond simple structural effects for two major reasons. First, the Bragg wavelengths along all three directions are almost identical (indicated by the yellow dashed line in Fig. 2(e)). Second, we performed challenging QPI measurements at temperatures up to 80 K and found that the difference between wavevectors, as defined by $\Delta k_1 \equiv [q_1^b - (q_1^a + q_1^c)/2]/2$ near the Fermi level, exhibits a strong temperature dependence (inset in Fig. 2(f) and Fig. 1(c)). These observations suggest that the anisotropic dispersion most likely originates from a new $q = 0$ nematic state residing in the in-plane V-$d$ orbitals, which is similar to the orbital-selective electronic nematicity observed in the isostructural titanium-based kagome metal $A$Ti$_3$Bi$_5$ (53–55).

To investigate the relationship between this new $q = 0$ nematic state and the rotational-symmetry-breaking CDW order, we move to the doping level of x = 0.12 (T$_{CDW}$ ≈ 50 K; Fig. S1; 56) beyond which the long-range CDW order vanishes. To examine the in-plane $2 \times 2$ CDW modulation, we acquired a series of bias-dependent topographies; two representative examples are shown in Figs. 3(a) and 3(b). At low energies, the presence of CDW order is evident in both real-space images and their FTs (Fig. 3(c)), which is C$_2$ symmetric with stronger modulation along one lattice direction. This anisotropy is reflected in the high intensity of the CDW peak in the FT linecuts shown in Fig. 3(d) (to eliminate tip anisotropy, the data are normalized by the corresponding Bragg peak intensities), and is also discernible in the real-space images (Fig. 3(b)). We note that the additional $4 \times 1$ charge order is suppressed at x = 0.12; thus, the observed anisotropy in CDW modulation signifies the emergence of a nematic $2 \times 2$ CDW order, as widely reported in this family of materials (39–42).

The measured QPI patterns at x = 0.12 closely resemble those observed at x=0.18. As shown in Figs. 3(e) to (h), the scattering vectors associated with Sb-derived states remain isotropic, whereas the in-plane V-derived scattering features retain a pronounced twofold anisotropy. Their energy dispersions are plotted in Fig. 3(h). Owing to the three-dimensional nature of the CDW order, the Sb $p_z$ band undergoes further folding and splits into two branches near $-65$ mV, yet the corresponding wavevectors remain essentially isotropic. In contrast, the dispersions of the V-derived scattering vectors exhibit pronounced anisotropy, satisfying $q_1^b > q_1^a \approx q_1^c$ and $q_2^b > q_2^a \approx q_2^c$. Surprisingly, the C$_2$ axis of the QPI patterns aligns along a different lattice direction (blue dashed arrow; direction B in Figs. 3(f) and (g)) compared to that of the nematic CDW (red dashed arrow; direction A). This misalignment, together with the emergence of the $q = 0$ state even in the absence of the CDW order, strongly suggests that these are two independent electronic states. Furthermore, the misalignment of their C$_2$ axes is not coincidental but rather a general feature at x=0.12, which has been consistently observed in different regions and different samples (Fig. S3–4; 56).

Next, we proceed to examine how these two electronic orders manifest in the pristine sample. As shown in Fig. 4(a), the QPI patterns near the Fermi level are also strikingly C$_2$ symmetric for x = 0. By tracking the energy dispersions, we find that the anisotropy in **q**$_1$ observed in the doped samples persists in the pristine samples (Fig. 4(b)). Such anisotropy cannot be readily explained by unidirectional electron–phonon coupling alone (42), as an electron-boson coupling typically alters the Fermi velocity while preserving the Fermi vector (i.e., the filling level). Notably, the C$_2$ axis of the QPI patterns is now aligned with the principal axis of the nematic $2 \times 2$ CDW order (Figs. 4(a) and 4(c)). These observations suggest that the $q = 0$ Fermi pocket distortion is already present at x = 0, with its principal axis pinned by the C$_2$ axis of the strong CDW order.

Given that a Ti substitution on the kagome network has a $C_2$ local crystalline environment (depending on the three sublattices labeled A, B, and C; Fig. 4(d)), it is essential to assess whether the local distortion around impurities influences the $C_2$-symmetric $q = 0$ state in the doped sample. To this end, we performed a statistical analysis to examine the relationship between any potential preferred impurity orientation and the nematic director of the electronic state. The orientation of each impurity can be unambiguously determined from high-bias topographic images (Fig. 4(e)). We then analyzed the distribution of impurity orientations in field of views of 40 nm × 40 nm and compared it with the symmetry axis of the $C_2$ QPI pattern. As shown in Fig. 4(f), no discernible correlation is found between the orientation distribution of Ti impurities and the $C_2$ axis of the electronic state, across different samples and spatial regions. These findings demonstrate that the observed $C_2$ electronic state is not a simple consequence of local impurity configurations, but rather arises from an intrinsic electronic instability of the system.

Our experiments reveal the emergence of two distinct types of nematic orders in Ti-doped $CsV_3Sb_5$. The first is tied to the in-plane 2×2 CDW, intimately related to the $d_{yz}$-orbital van Hove singularity with pronounced Fermi surface nesting (18,19). This order develops simultaneously with the CDW and can be expressed as $\vec{\phi} = 1/\sqrt{6}(M_2^2 + M_3^2 - 2M_1^2, \sqrt{3}(M_2^2 - M_3^2))$, where $M_i$ are the three CDW components. The other is an intrinsic $q = 0$ electronic nematicity parametrized as $\vec{\eta} = (\eta_1, \eta_2)$, which features a nematic distortion on the $d_{x^2-y^2}$-orbital dominated triangular pockets around the K point and could be a nematic bond order that stems from high-order van Hove singularity (27). With the orientation of CDW-induced nematicity set to be the $M_1$ axis, $M = (\Delta_0 + \delta, \Delta_0, \Delta_0)$, the interplay between these orders is captured by a Landau free energy $f = f_\phi + f_\eta + f_{\phi-\eta}$. The two individual terms are $f_\phi \approx \alpha_\phi \phi + \beta_\phi \phi^2$ and $f_\eta = \alpha_\eta \eta^2 + \beta_\eta \eta^3 cos6\theta_\eta + \gamma_\eta \eta^4$ with $\phi = |\vec{\phi}|$, $\eta = |\vec{\eta}|$ and $\theta_\eta$ being the nematic angle, containing both quadratic and cubic terms that determine the orientation of intrinsic nematicity. The bilinear coupling term $f_{\phi-\eta} = \beta_{\phi\eta}\vec{\phi} \cdot \vec{\eta}$ dictates their relative alignment. This theoretical model naturally explains the experimentally observed doping evolution of CDW-induced and intrinsic nematicities and their relative orientation, as summarized in Fig. 5. At high Ti doping (x=0.18), the $2 \times 2$ CDW is completely suppressed ($\alpha_\phi, \beta_{\eta\phi} > 0$, $\phi \to 0$), leaving only the $q = 0$ nematicity. At intermediate doping (x=0.12), both orders coexist ($\alpha_\phi < 0$), but a weak positive inter-order coupling $\beta_{\phi\eta}$ results in misaligned nematic orientations. Approaching the pristine compound, the CDW predominates and the coupling $\beta_{\phi\eta}$ changes sign. This negative coupling generates an energetically locked state with aligned nematic orders. The whole evolution can be tracked by the dashed arrows in the calculated phase diagram as shown in Fig. 5 with $\alpha_\eta < 0$ (see more details in 56; SI note 6). Microscopically, this evolving coupling is likely driven by doping-induced changes in the hybridization between the $d_{yz}$ and $d_{x^2-y^2}$ orbitals that dictate the two nematic instabilities.

Our results provide valuable insight into the understanding of unusual correlated phenomena in these materials. First, the discovery of an additional $q = 0$ state indicates that the electronic nematicity observed in this family is not merely a consequence of the interlayer stacking order of the CDW. Instead, this $q = 0$ nematic state appears to be independent, since it persists even when the long-range CDW phase is totally suppressed, and its principal axis can be misaligned with that of the nematic CDW. Moreover, we found that $\Delta k_1$, a measure of the distortion strength of the corresponding Fermi pockets, exhibits a dome-shaped dependence on doping, in contrast to the monotonic suppression of the CDW order (Fig.S9; 56). Second, temperature-dependent QPI measurements reveal that the $q=0$ nematic state remains detectable up to high temperatures in doped samples (Fig. 1c). This is an important observation, as in the pristine sample the QPI patterns associated with the V-d-orbitals vanish above T* ≈ 35 K. This striking discrepancy points to a more complex microscopic interplay between the nematic CDW and the $q=0$ order beyond the phenomenological Ginzburg-Landau description, where significant electronic fluctuation appears above T* before these two orders are locked by the lattice. Interestingly, sharp quasi-one-dimensional scattering features arising from out-of-plane V-$d_{yz}$ orbitals emerge below T* in the pristine compound, which is in line with the coherent out-of-plane charge transport (57). Given that the nematic axes of the two orders are not always aligned, it would be intriguing to explore whether their interplay could give rise to electronic chirality in these systems. Finally, this work underscores the intrinsic propensity toward electronic nematicity on the kagome lattice and establishes a material platform in which two distinct types of nematic orders develop and intertwine.

Note added: During the preparation of this manuscript, we became aware of two related STM works (58,59) on doped $CsV_3Sb_5$.

# References

[1] E. Fradkin, S. A. Kivelson, and J. M. Tranquada, Colloquium: Theory of intertwined orders in high-temperature superconductors, Rev. Mod. Phys. 87, 457 (2015).
[2] R. M. Fernandes, P. P. Orth, and J. Schmalian, Intertwined vestigial order in quantum materials: Nematicity and beyond, Annu. Rev. Condens. Matter Phys. 10, 133 (2019).
[3] D. F. Agterberg, J. C. S. Davis, S. D. Edkins, E. Fradkin, D. J. Van Harlingen, S. A. Kivelson, P. A. Lee, L. Radzihovsky, J. M. Tranquada, and Y. Wang, The physics of pair-density waves: Cuprate superconductors and beyond, Annu. Rev. Condens. Matter Phys. 11, 231 (2020).
[4] L. Balents, Spin liquids in frustrated magnets, Nature 464, 199 (2010).
[5] W.-H. Ko, P. A. Lee, and X.-G. Wen, Doped kagome system as exotic superconductor, Phys. Rev. B 79, 214502 (2009).
[6] R. Nandkishore, L. S. Levitov, and A. V. Chubukov, Chiral superconductivity from repulsive interactions in doped graphene, Nat. Phys. 8, 158 (2012).
[7] I. Syôzi, Statistics of kagomé lattice, Prog. Theor. Phys. 6, 306 (1951).
[8] J.-X. Yin, B. Lian, and M. Z. Hasan, Topological kagome magnets and superconductors, Nature 612, 647 (2022).
[9] S.-L. Yu and J.-X. Li, Chiral superconducting phase and chiral spin-density-wave phase in a Hubbard model on the kagome lattice, Phys. Rev. B 85, 144402 (2012).

[10] M. L. Kiesel and R. Thomale, Sublattice interference in the kagome Hubbard model, Phys. Rev. B 86, 121105 (2012).

[11] M. L. Kiesel, C. Platt, and R. Thomale, Unconventional Fermi surface instabilities in the kagome Hubbard model, Phys. Rev. Lett. 110, 126405 (2013).

[12] W.-S. Wang, Z.-Z. Li, Y.-Y. Xiang, and Q.-H. Wang, Competing electronic orders on kagome lattices at van Hove filling, Phys. Rev. B 87, 115135 (2013).

[13] B. R. Ortiz et al., New kagome prototype materials: Discovery of $KV_3Sb_5$, $RbV_3Sb_5$, and $CsV_3Sb_5$, Phys. Rev. Materials 3, 094407 (2019).

[14] B. R. Ortiz et al., $CsV_3Sb_5$: A $Z_2$ topological kagome metal with a superconducting ground state, Phys. Rev. Lett. 125, 247002 (2020).

[15] T. Neupert, M. M. Denner, J.-X. Yin, R. Thomale, and M. Z. Hasan, Charge order and superconductivity in kagome materials, Nat. Phys. 18, 137 (2022).

[16] K. Jiang, T. Wu, J.-X. Yin, Z. Wang, M. Z. Hasan, S. D. Wilson, X. Chen, and J. Hu, Kagome superconductors $AV_3Sb_5$ (A = K, Rb, Cs), Natl. Sci. Rev. 10, nwac199 (2023).

[17] S. D. Wilson and B. R. Ortiz, $AV_3Sb_5$ kagome superconductors, Nat. Rev. Mater. 9, 420 (2024).

[18] M. Kang et al., Twofold van Hove singularity and origin of charge order in topological kagome superconductor $CsV_3Sb_5$, Nat. Phys. 18, 301 (2022).

[19] Y. Hu et al., Rich nature of van Hove singularities in kagome superconductor $CsV_3Sb_5$, Nat. Commun. 13, 2220 (2022).

[20] Y. Hu, X. Wu, A. P. Schnyder, and M. Shi, Electronic landscape of kagome superconductors $AV_3Sb_5$ (A = K, Rb, Cs) from angle-resolved photoemission spectroscopy, npj Quantum Mater. 8, 67 (2023).

[21] M. H. Christensen, T. Birol, B. M. Andersen, and R. M. Fernandes, Theory of the charge density wave in $AV_3Sb_5$ kagome metals, Phys. Rev. B 104, 214513 (2021).

[22] M. M. Denner, R. Thomale, and T. Neupert, Analysis of charge order in the kagome metal $AV_3Sb_5$ (A = K, Rb, Cs), Phys. Rev. Lett. 127, 217601 (2021).

[23] X. Wu et al., Nature of unconventional pairing in the kagome superconductors $AV_3Sb_5$ (A = K, Rb, Cs), Phys. Rev. Lett. 127, 177001 (2021).

[24] S. Zhou and Z. Wang, Chern Fermi pocket, topological pair-density wave, and charge-4e and charge-6e superconductivity in kagome superconductors, Nat. Commun. 13, 7288 (2022).

[25] R. Tazai, Y. Yamakawa, and H. Kontani, Charge-loop current order and $Z_3$ nematicity mediated by bond-order fluctuations in kagome metals, Nat. Commun. 14, 7845 (2023).

[26] H. Li, Y. B. Kim, and H.-Y. Kee, Intertwined van Hove singularities as a mechanism for loop current order in kagome metals, Phys. Rev. Lett. 132, 146501 (2024).

[27] X. Han, A. P. Schnyder, and X. Wu, Enhanced nematicity emerging from higher-order Van Hove singularities, Phys. Rev. B 107, 184504 (2023).

[28] R. Fu, J. Zhan, M. Dürrnagel, H. Hohmann, R. Thomale, J. Hu, Z. Wang, S. Zhou, and X. Wu, Exotic charge-density waves and superconductivity on the kagome lattice, Natl. Sci. Rev. **12**, nwaf414 (2025).

[29] J. Zhan, H. Hohmann, M. Dürrnagel, R. Fu, S. Zhou, Z. Wang, R. Thomale, X. Wu, and J. Hu, Loop current order on the kagome lattice, Phys. Rev. Lett. 136, 126001 (2026).

[30] Y.-X. Jiang et al., Unconventional chiral charge order in kagome superconductor $KV_3Sb_5$, Nat. Mater. 20, 1353 (2021).

[31] C. Mielke III et al., Time-reversal symmetry-breaking charge order in a kagome superconductor, Nature 602, 245 (2022).
[32] C. Guo et al., Switchable chiral transport in charge-ordered kagome metal $CsV_3Sb_5$, Nature 611, 461 (2022).
[33] Y. Xu, Z. Ni, Y. Liu, B. R. Ortiz, Q. Deng, S. D. Wilson, B. Yan, L. Balents, and L. Wu, Three-state nematicity and magneto-optical Kerr effect in the charge density waves in kagome superconductors, Nat. Phys. 18, 1470 (2022).
[34] Y. Xing et al., Optical manipulation of the charge-density-wave state in $RbV_3Sb_5$, Nature 631, 60 (2024).
[35] Y. Hu et al., Time-reversal symmetry breaking in charge density wave of $CsV_3Sb_5$ detected by polar Kerr effect, arXiv:2208.08036 (2022).
[36] H. Gui et al., Probing orbital magnetism of a kagome metal $CsV_3Sb_5$ by a tuning fork resonator, Nat. Commun. 16, 4275 (2025).
[37] H. Zhao, H. Li, B. R. Ortiz, S. M. Teicher, T. Park, M. Ye, Z. Wang, L. Balents, S. D. Wilson, and I. Zeljkovic, Cascade of correlated electron states in the kagome superconductor $CsV_3Sb_5$, Nature 599, 216 (2021).
[38] Y. Xiang, Q. Li, Y. Li, W. Xie, H. Yang, Z. Wang, Y. Yao, and H. H. Wen, Twofold symmetry of c-axis resistivity in topological kagome superconductor $CsV_3Sb_5$ with in-plane rotating magnetic field, Nat. Commun. 12, 6727 (2021).
[39] L. Nie et al., Charge-density-wave-driven electronic nematicity in a kagome superconductor, Nature 604, 59 (2022).
[40] H. Li, H. Zhao, B. R. Ortiz, T. Park, M. Ye, L. Balents, Z. Wang, S. D. Wilson, and I. Zeljkovic, Rotation symmetry breaking in the normal state of a kagome superconductor $KV_3Sb_5$, Nat. Phys. 18, 265 (2022).
[41] H. Li, H. Zhao, B. R. Ortiz, Y. Oey, Z. Wang, S. D. Wilson, and I. Zeljkovic, Unidirectional coherent quasiparticles in the high-temperature rotational symmetry broken phase of $AV_3Sb_5$ kagome superconductors, Nat. Phys. 19, 637 (2023).
[42] P. Wu et al., Unidirectional electron–phonon coupling in the nematic state of a kagome superconductor, Nat. Phys. 19, 1143 (2023).
[43] H. Chen et al., Roton pair density wave in a strong-coupling kagome superconductor, Nature 599, 222 (2021).
[44] H. Deng et al., Chiral kagome superconductivity modulations with residual Fermi arcs, Nature 632, 775 (2024).
[45] D. Chen et al., Anomalous thermoelectric effects and quantum oscillations in the kagome metal $CsV_3Sb_5$, Phys. Rev. B 105, L201109 (2022).
[46] C. Guo et al., Correlated order at the tipping point in the kagome metal $CsV_3Sb_5$, Nat. Phys. 20, 579 (2024).
[47] C. Guo et al., Many-body interference in kagome crystals, Nature 647, 68 (2025).
[48] H. Li et al., Small Fermi pockets intertwined with charge stripes and pair density wave order in a kagome superconductor, Phys. Rev. X 13, 031030 (2023).
[49] H. Yang et al., Titanium doped kagome superconductor $CsV_{3-x}Ti_xSb_5$ and two distinct phases, Sci. Bull. 67, 2176 (2022).
[50] Y. Liu et al., Doping evolution of superconductivity, charge order, and band topology in hole-doped topological kagome superconductors $Cs(V_{1-x}Ti_x)_3Sb_5$, Phys. Rev. Materials 7, 064801 (2023).
[51] Z. Wu et al., Competitive charge density waves in the doped kagome superconductor $CsV_{3-x}Ti_xSb_5$, Phys. Rev. B 112, 144512 (2025).

[52] Z. Huang et al., Revealing the orbital origins of exotic electronic states with Ti substitution in kagome superconductor $CsV_3Sb_5$, Phys. Rev. Lett. 134, 056001 (2025).
[53] H. Li et al., Electronic nematicity without charge density waves in titanium-based kagome metal, Nat. Phys. 19, 1591 (2023).
[54] H. Yang et al., Superconductivity and nematic order in a new titanium-based kagome metal $CsTi_3Bi_5$ without charge density wave order, Nat. Commun. 15, 9626 (2024).
[55] Y. Hu et al., Non-trivial band topology and orbital-selective electronic nematicity in a titanium-based kagome superconductor, Nat. Phys. 19, 1827 (2023).
[56] See Supplemental Material at : for transport characterization, a table of Ti concentration by counting, more QPI data for x=0.12 and 0.18, more data showing misalignment between the nematic axes of the CDW and the q=0 nematic state in x=0.12 samples, QPI linecuts for x=0.12 and 0.18, a Ginzburg-Landau analysis of intertwined CDW and electronic nematic order, and the doping dependence of the Fermi pocket distortion.
[57] C. Guo et al., Many-body interference in kagome crystals, Nature 647, 68 (2025).
[58] Z. Huang et al., The parent state in kagome metals and superconductors: Chiral-nematic Fermi liquid state, arXiv:2511.09402 (2025).
[59] M. Xu et al., Pervasive electronic nematicity as the parent state of kagome superconductors, arXiv:2511.22002 (2025).

**Acknowledgements**

We thank Yingying Peng and Hui Chen for valuable discussions. This work is supported by the National Natural Science Foundation of China (Grant Nos. 52261135638, 12488201, 52373309), the Quantum Science and Technology-National Science and Technology Major Project (Grant No. 2021ZD0302800), the National Key R&D Program of the MOST of China (Grants No. 2022YFA1602600), the Scientific Research Innovation Capability Support Project for Young Faculty (No. ZYGXQNJSKYCXNLZCXM-M25), and the HFNL Self-Deployed Project (ZB2025020200, ZB2025020100).

**Figure 1**

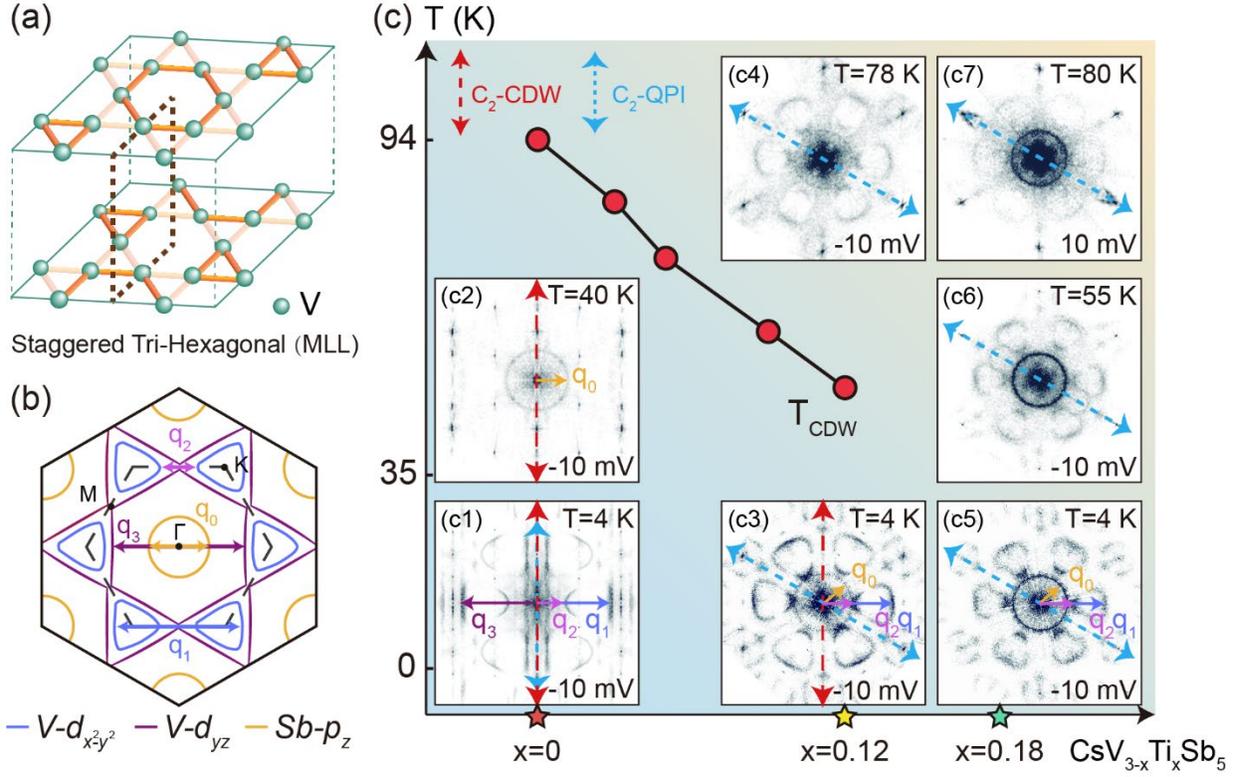

FIG. 1. Evolution of QPI patterns in CsV$_3$Sb$_5$ as a function of Ti doping and temperature. (a) Staggered tri-hexagonal configuration of the 2×2×2 CDW with a $\pi$ phase shift between adjacent layers. (b) Schematic of the constant-energy contours (CECs) near E$_F$. The wavevectors **q$_i$** (i=0, 1, 2, 3) represent the dominant scattering channels. (c) The evolution of QPI as a function of Ti doping and temperature. With increasing Ti concentration, the 2×2×2 CDW is progressively suppressed (red dots). The blue and red double-headed arrows indicate the C$_2$ axis of the QPI patterns and the in-plane CDW modulations.

**Figure 2**

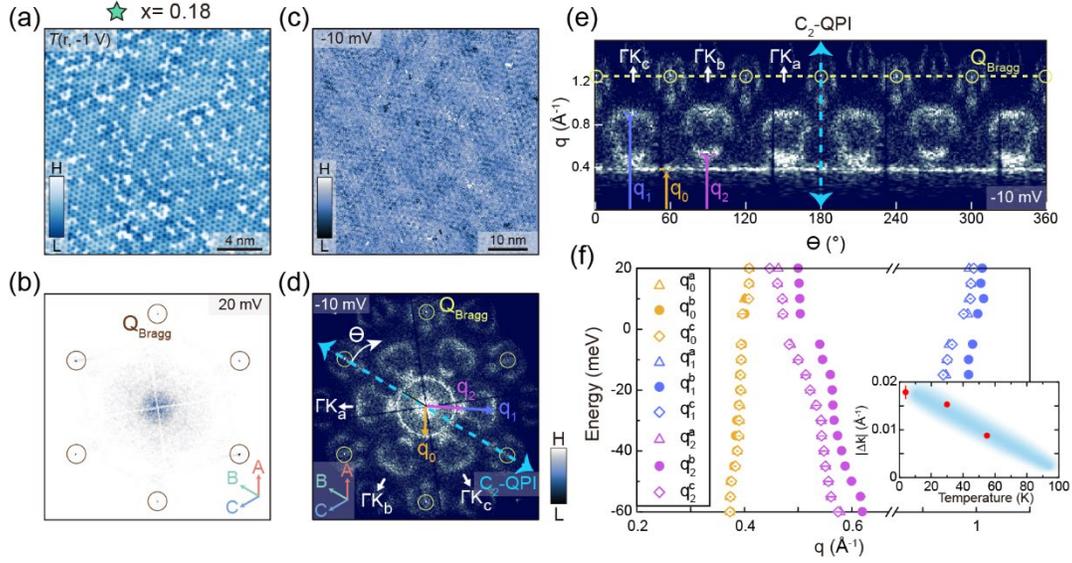

FIG. 2. Observation of a $q = 0$ nematic state at x = 0.18. (a) STM topography of the Sb-terminated surface. The bright protrusions correspond to Ti atoms that replace V atoms in the underneath kagome layer. (b) FFT of the Sb-terminated surface topography acquired at 20 mV, showing only the Bragg peaks. (c, d) Differential conductance map $L(\mathbf{r}, -10\ \mathrm{mV})$ and its Fourier transform. A, B, and C indicate the three Bragg directions. The blue double-headed arrows mark the $C_2$ symmetry axis of the QPI pattern. (e) Angular-dependent QPI intensity map of (d). (f) Energy dispersions of the QPI scattering vectors, extracted along the three $\Gamma K$ directions. The fitting procedure is shown in Fig.S5 (56). Inset: Temperature-dependent $\Delta k(\boldsymbol{q_1})$. The QPI images are raw data without symmetrization.

# Figure 3

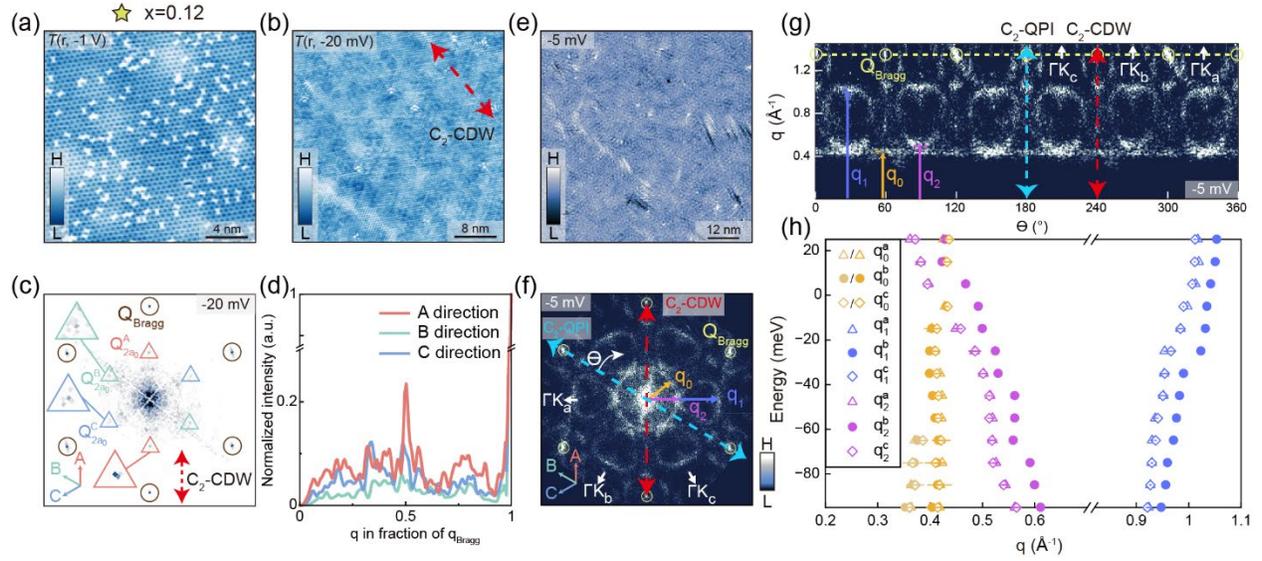

FIG. 3. Coexistence of the nematic CDW and the $q = 0$ nematic state for x=0.12. (a) Topographic image of the Sb-terminated surface. (b) Topography recorded at -20 mV revealing the $2 \times 2$ charge modulation. (c) Fourier transform of (b), showing the $2 \times 2$ CDW peaks (triangles) in addition to the Bragg peaks (circles). (d) FT line cuts taken along the three atomic directions in (c), in which a pronounced anisotropy of the CDW peak can be found. (e, f) Differential conductance map $L(\mathbf{r}, -5\text{ mV})$ and its FT. The blue and red arrows mark the $C_2$ symmetry axes of the QPI pattern and CDW modulation, respectively. (g) Angular-dependent QPI intensity map of (f). (h) Energy dispersions of the QPI scattering vectors, extracted along the three $\Gamma K$ directions. The fitting procedure is shown in Fig.S5 (56).

**Figure 4**

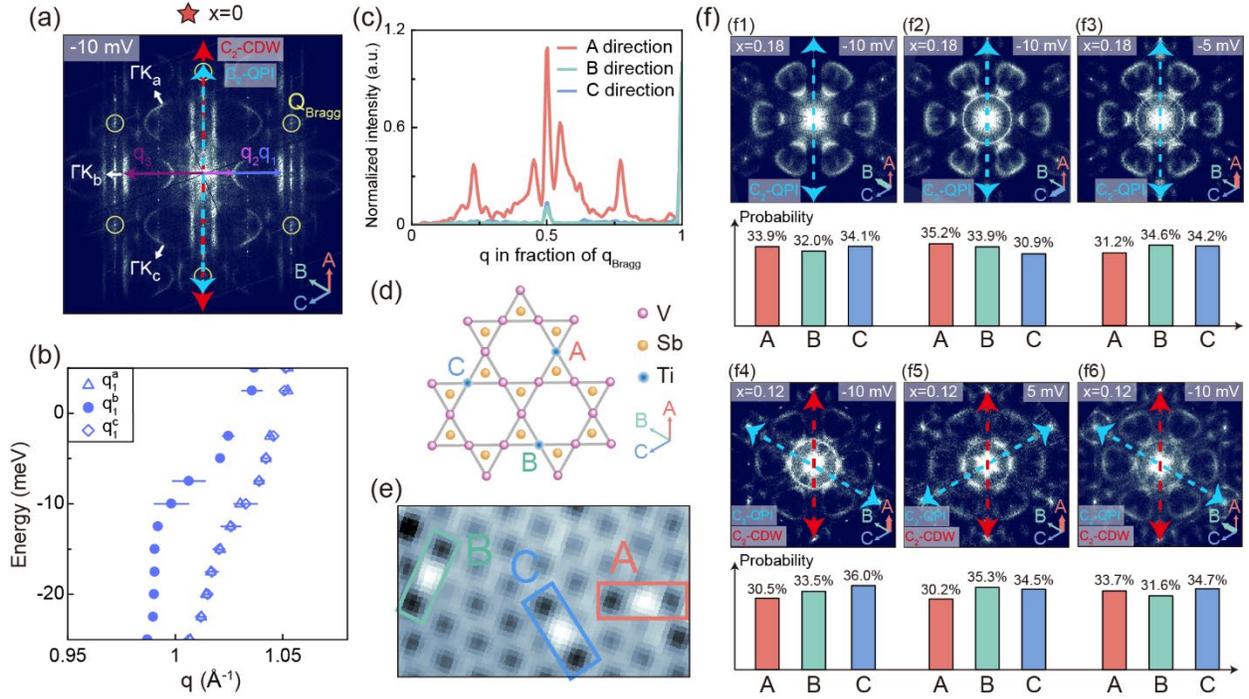

FIG. 4. Electronic orders in pristine $CsV_3Sb_5$ and statistical analysis of the dopant-induced anisotropy. (a) FT of $L(\mathbf{r}, -10\ mV)$ for the pristine sample which exhibits prominent unidirectional features. (b) Energy dispersions of the QPI scattering vectors, extracted along the three $\Gamma K$ directions. (c) FT intensity profiles along the three atomic directions of the corresponding topographic image of (a), showing nearly twofold-symmetric CDW modulation. The $C_2$ symmetry axes of the QPI pattern and CDW modulation are aligned as shown in (a). (d) Three types of atomic positions for the Ti dopants on the kagome lattice. (e) STM topographic image showing three Ti dopants at the corresponding sublattice sites depicted in (d). (f) QPI patterns recorded in different regions of different x=0.18 and x=0.12 samples. The lower panels display the statistical analysis ( in FOVs of typical 40 nm × 40 nm) for the numbers of Ti impurities located on these three sublattices. No discernable correlation between the impurity orientation and the $C_2$ axis of the QPI patterns is found.

**Figure 5**

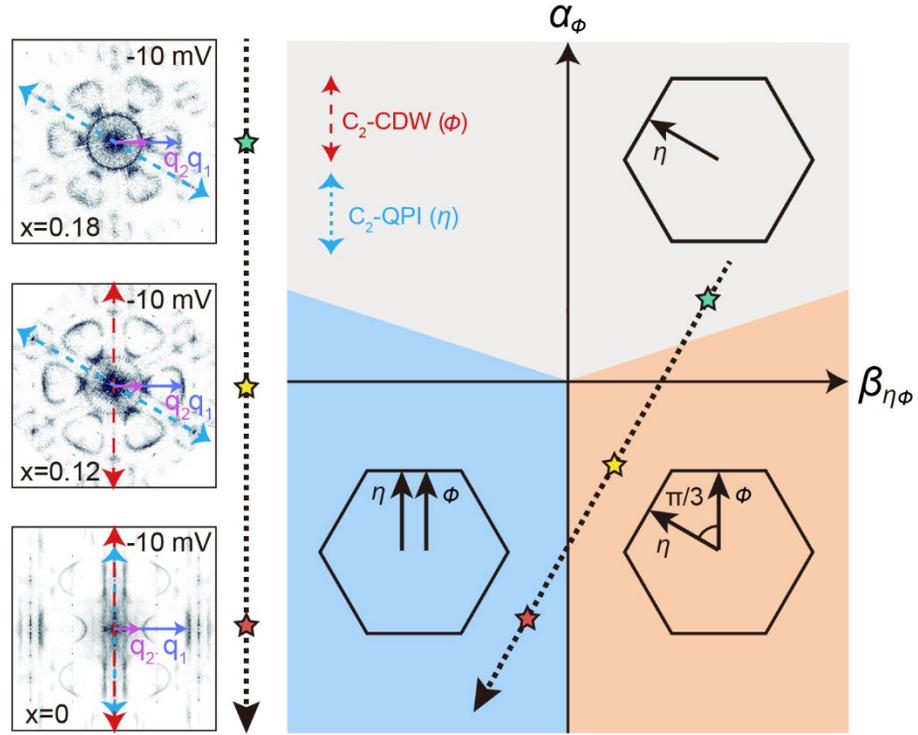

FIG. 5. Ginzburg-Landau phase diagram of the coupled CDW and nematic orders in the $\alpha_\phi - \beta_{\eta\phi}$ plane. Here $\alpha_\phi$ is the quadratic coefficient of the nematic CDW component, and $\beta_{\eta\phi}$ characterizes the coupling between the CDW-induced nematic order $\phi$ and the intrinsic nematic order $\eta$. Different colors indicate the three phases: a coexistence phase with aligned nematic directors for $\beta_{\eta\phi} < 0$, a coexistence phase with a relative rotation of $\pi/3$ for $\beta_{\eta\phi} > 0$, and a pure intrinsic nematic phase in which the CDW is suppressed. The black dashed arrow marks a representative evolution path with decreasing doping. Insets illustrate the corresponding ground-state configurations, where $\phi$ and $\eta$ denote the CDW-induced and intrinsic nematic orders, respectively. The left panel shows the evolution of the QPI patterns with doping.